\documentstyle[prd,aps,floats]{revtex}

\begin{document}
\preprint{astro-ph/0010082}
\draft

\input epsf

\renewcommand{\topfraction}{0.99}
\renewcommand{\bottomfraction}{0.99}

\twocolumn[\hsize\textwidth\columnwidth\hsize\csname
@twocolumnfalse\endcsname

\title{Inflationary perturbations near horizon crossing}
\author{Samuel M.~Leach and Andrew R.~Liddle}
\address{Astronomy Centre,
University of Sussex, Brighton BN1 9QJ, United Kingdom} 
\date{\today} 
\maketitle
\begin{abstract}
We study the behaviour of inflationary density perturbations in the
vicinity of horizon crossing, using numerical evolution of the
relevant mode equations. We explore two
specific scenarios. In one, inflation is temporarily ended
because a portion of the potential is too steep to support inflation.
We find that perturbations on super-horizon scales can be 
modified, usually leading to a large amplification, because of entropy 
perturbations in the scalar field. This leads to a broad feature in the power
spectrum, and the slow-roll and Stewart--Lyth approximations, which assume 
the perturbations reach an asymptotic regime well outside the horizon, can
fail by many orders of magnitude in this regime. In the second
scenario we consider perturbations generated right at the end of
inflation, which re-enter shortly after inflation ends --- such
perturbations can be relevant for primordial black hole formation.
\end{abstract}

\pacs{PACS numbers: 98.80.Cq \hfill astro-ph/0010082}

\vskip2pc]

\section{Introduction}

The inflationary cosmology remains the leading candidate theory for the origin
of structure (see Refs.~\cite{LLrep,LLbook} for reviews).  Considerable 
attention has
been focussed on producing highly accurate calculations of the perturbations
produced during inflation \cite{perts}, which ultimately may be measured at the
percent level via the microwave anisotropies they induce \cite{measure}.
Special interest has been given to the simplest case, where there is only a
single scalar field degree of freedom during inflation, and we will focus
exclusively on that case in this paper.

In considering how a perturbation might evolve, a crucial quantity is
the comparison of the inverse wavenumber with the Hubble length, given
by the ratio $aH/k$. By definition inflation corresponds to a
decreasing comoving Hubble length, $d(aH)/dt >0$; consequently when
modes of a given wavenumber are considered, they begin their evolution
well inside the horizon\footnote{We use `horizon' as shorthand for the
more precise `Hubble length'; it does not refer to the particle
horizon.} and cross outside during inflation. At early times, flat
space-time quantum field theory can be used to fix the initial
normalization of the perturbations. Around the epoch of horizon
crossing, the perturbation becomes frozen in, corresponding to a
constant curvature perturbation once the mode is well outside the
horizon. This leads to standard expressions for the spectrum of
density perturbations such as
\begin{equation}
\label{eqn:scalpert}
{\cal P}_{{\cal R}}^{1/2}(k) = \left. \frac{1}{2\pi} \,
	\frac{H^2}{|\dot{\phi}|} \right|_{k=aH} \,,
\end{equation}
where ${\cal P}_{{\cal R}}$ is the power spectrum of the curvature
perturbation ${\cal R}$, using the notation of Refs.~\cite{LLrep,LLbook}.  This
expression gives the amplitude of the perturbations in terms of the
values of the Hubble parameter $H$ and scalar field velocity
$\dot{\phi}$ at the time the mode crossed the horizon, i.e.~when
$k=aH$. A similar expression holds for the gravitational wave
amplitude, and more sophisticated higher-order versions have also been
derived.

Eq.~(\ref{eqn:scalpert}) gives the impression that the amplitude of
perturbations is being quoted at the instant of horizon
crossing. However, it is important to realize that that is not in
fact the case. The value quoted is the perturbation amplitude attained
in the asymptotic limit $k/aH \rightarrow 0$ when the perturbation is
well outside the horizon; it just happens to be written in terms of
the values the background parameters had at the instant of horizon
crossing. In fact the value of the curvature perturbation at the
instant of horizon crossing typically differs by a significant factor
from the asymptotic value.

Bearing that in mind, in this paper we consider two circumstances
where the standard formula for the perturbation amplitude may not be
valid, due to a failure to reach the asymptotic limit. In each case,
this is because inflation ends before the true asymptotic regime has
been reached. In our first scenario, we consider a temporary
interruption to inflation, where the field driving inflation has a
region where the potential is too steep to sustain inflation. We will
see that this can lead to significant modifications to the standard
results, and indeed that even modes significantly outside the horizon
can receive a large change in amplitude due to entropy perturbations 
in the scalar field. In the second scenario,
we consider perturbations produced at the end of inflation, where
their amplitude on re-entry has relevance to the production of
primordial black holes.

\section{Formalism}

The scalar perturbations are best followed using the variable $u = a \,
\delta \phi$ \cite{Muk,MFB}, and the equation satisfied by its Fourier
modes $u_{{\rm k}}$ is 
\begin{equation}
u_{{\rm k}}''+\left(k^2-\frac{z''}{z}\right)u_{{\rm k}}=0 \,,
\label{eqn:mode1}
\end{equation}
where primes denote differentiation with respect to conformal time, 
$z\equiv a\dot{\phi}/H$, and 
\begin{equation}
\frac{z''}{z}=2a^2H^2 \left[ 1+\epsilon-\frac{3}{2}\eta+\epsilon^2 
	-2\epsilon\eta +\frac{1}{2}\eta^2 + \frac{1}{2} \xi^2 \right], 
\label{eqn:d2zdt2}
\end{equation}
where we define the Hubble slow-roll parameters \cite{LPB} as
\begin{eqnarray}
\label{eqn:slow-roll_hubble}
\epsilon &\equiv& \frac{m_{{\rm Pl}}^2}{4\pi}\left(\frac{H_{,\phi}}{H}\right)^2
    =  3 \, \frac{\dot{\phi}^2/2}{V+\dot{\phi}^2/2} \,\,;\\
 \eta &\equiv& \frac{m_{{\rm Pl}}^2}{4\pi} \, \frac{H_{,\phi\phi}}{H}
      = -3 \, \frac{\ddot{\phi}}{3H\dot{\phi}} \,\,; \\ 
 \xi^2 &\equiv& \frac{m^4_{{\rm Pl}}}{16\pi^2}\frac{H_{,\phi}\,H_{,\phi \phi 
\phi}}{H^2}
     = 3(\epsilon+\eta)-\eta^2-\frac{V_{,\phi\phi}}{H^2} \,,
\label{eqn:xi2}
\end{eqnarray}
where `${}_{,\phi}$' denotes differentiation with respect to $\phi$.
Mode equation~(\ref{eqn:mode1}) has two asymptotic regimes
characterized by the relative sizes of $k^2$ and $z''/z$. Only in the
slow-roll limit, where $\epsilon$,$|\eta|$,$|\xi^2| \ll 1$, will this
necessarily be the same as comparing $k$ and $aH$. Bearing this in
mind, well within the horizon in the limit of $aH/k\rightarrow 0$ each
mode behaves like a free field
\begin{equation}
u_{{\rm k}} \rightarrow \frac{1}{\sqrt{2k}} \, e^{-ik\tau}\,,
\label{eqn:freefield}
\end{equation}
while in the limit $k^2 \ll z''/z$ we have a growing mode solution
\begin{equation}
u_{{\rm k}} \propto z\,,
\label{eqn:regmode}
\end{equation}
which means that the curvature perturbation, $|{\mathcal R}_{{\rm
k}}|=|u_{{\rm k}}/z|$, remains constant on superhorizon scales. The quantity $z$ 
is sometimes described as the pump field for scalar perturbations. 

To calculate the constant of proportionality of Eq.~($\ref{eqn:regmode}$)
it is generally assumed that $\epsilon$ and $\eta$ are slowly varying
at around horizon crossing, which will be a valid approximation as
long as these parameters are small, since
\begin{equation}
\frac{\epsilon'}{aH}=2\epsilon(\epsilon-\eta) \,\, ; \,\,
  \frac{\eta'}{aH}= \epsilon\eta-\xi^2 \,.
\label{eqn:epsilon_dot}
\end{equation}
The expression for conformal time $\tau$ then takes on a simple form
\begin{equation}
\tau \equiv \int\frac{dt}{a} \simeq - \frac{1}{aH} \, \frac{1}{1-\epsilon} \,,
\label{eqn:conformal_t}
\end{equation}
and Eq.~(\ref{eqn:mode1}) reduces to a Bessel equation leading to the
Stewart--Lyth result for the power spectrum \cite{SL}
\begin{equation}
{\mathcal P}_{{\mathcal R}}^{1/2}\simeq\left[1-(2C+1)\epsilon-C\eta\right]
	\frac{1}{2\pi}\left.\frac{H^2}{|\dot{\phi}|}\right|_{k=aH} \,,
\label{eqn:SL}
\end{equation}
where $C \simeq -0.73$ is a numerical constant.

In fact, the constancy of ${\mathcal R}_{{\rm k}}$ depends on
$\epsilon$ and $\eta$ doing nothing dramatic even \emph{after} horizon
crossing, which can be seen if we rewrite Eq.~(\ref{eqn:mode1}) in
terms of ${\mathcal R}_{{\rm k}}$ itself
\begin{equation}
{\mathcal R}_{{\rm k}}'' + 2\frac{z'}{z}{\mathcal R}_{{\rm k}}'+
	k^2{\mathcal R}_{{\rm k}}=0 \,,
\label{eqn:mode2}
\end{equation}
where
\begin{equation}
\frac{z'}{z}=aH \left[ 1+\epsilon-\eta \right] \,.
\label{eqn:dzdt}
\end{equation}
Initially ${\mathcal R}_{{\rm k}}$, like $u_{{\rm k}}$, will be
oscillating, although we are only interested in the envelope of this
oscillation, $\left|{\mathcal R}_{{\rm k}}\right|$, because only the
envelope contributes to the value of the real space curvature
perturbation, ${\mathcal R}$. Mode equation~(\ref{eqn:mode2}) is of
the form of a damped harmonic oscillator. In the slow-roll limit, at around 
horizon crossing
the system becomes dominated by the exponentially growing friction
term proportional to ${\mathcal R}_{{\rm k}}'$, and the solution to
Eq.~(\ref{eqn:mode2}) soon becomes well approximated by the form
\begin{equation}
\label{eqn:rk_soln}
\label{eqn:rdot_suppression} 
{\mathcal R}_{{\rm k}}(\tau) = {\rm const} \quad ; \quad 
\frac{{\mathcal R}_{{\rm k}}'(\tau)}{aH} \simeq {\rm const} \times 
\exp\left(-2N\right) \,,
\end{equation}
where $N$ is the number of
$e$-folds after horizon crossing.  The rapid freezing in of the curvature 
perturbation is apparent from Eq.~(\ref{eqn:rdot_suppression}), which measures 
the rate of change of ${\mathcal R}$ per Hubble time. We will examine the 
properties of the late-time solution in more detail later in the paper. 

In this paper we are interested in a more dramatic situation, arising 
through a failure of slow-roll. If at any time after horizon crossing the 
friction term in Eq.~(\ref{eqn:mode2}) \emph{changes sign} and becomes a 
negative driving term, then we can expect dramatic effects on modes which have 
recently left the horizon. This change of sign will occur whenever
$z$ reaches a local maximum, or equivalently whenever
\begin{equation}
1+\epsilon-\eta=0 \,.
\label{eqn:z-flip}
\end{equation}
As $\epsilon$ is always positive, $\eta$ must be at least one for this
to happen, which implies that a turn around in $z$ can occur only during a 
transition to fast-roll inflation or to a non-inflationary period. 

Before progressing to specific applications of the above, we note that
no such interesting effects can occur for gravitational waves. Their mode
equation is given by
\begin{equation}
v_{{\rm k}}''+\left(k^2-\frac{a''}{a}\right)v_{{\rm k}}=0 \,,
\label{eqn:mode_grav}
\end{equation}
where
\begin{equation}
\frac{a''}{a}=2a^2H^2\left[ 1-\frac{1}{2}\epsilon\right] \,.
\label{eqn:mode_grav2}
\end{equation}
In analogy with the above, Eq.~(\ref{eqn:mode_grav})
can be written as
\begin{equation}
V_{{\rm k}}'' + 2aHV_{{\rm k}}'+k^2V_{{\rm k}}=0 \,,
\label{eqn:mode_grav3}
\end{equation}
where
\begin{equation}
V_{{\rm k}} = \frac{v_{{\rm k}}}{a}\,.
\label{eqn:mode_grav4}
\end{equation}
The pump field $a$ increases for all time, and so the constancy
of the gravitational wave amplitude after horizon exit is assured,
until horizon re-entry.

\begin{figure}[t]
\centering 
\leavevmode\epsfysize=6cm \epsfbox{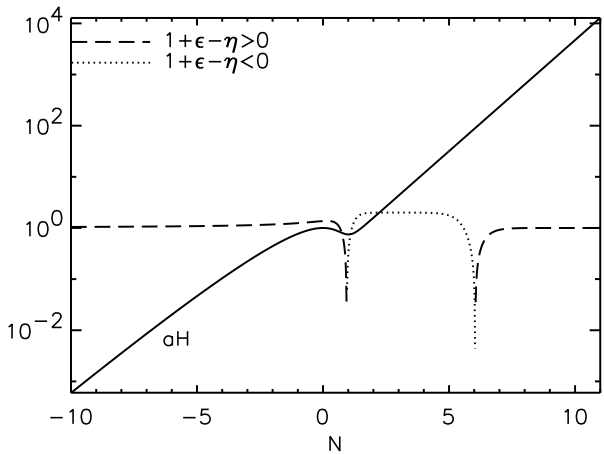}\\
\caption[fig1]{\label{fig:aH} The comoving Hubble wavenumber, $aH$,
increases with the number of $e$-folds of expansion, $N$, during
inflation. We set $N=0$ when inflation first ends, and inflation is
suspended for around 1 $e$-fold. The quantity $1+\epsilon-\eta$
remains negative for around 5 $e$-folds. We took $B=0.55$.}
\end{figure}

\section{Temporary interruption of inflation}

To illustrate the points made in the previous section, we study the specific 
example of
a false vacuum inflation model with a quartic potential~\cite{RLL}
\begin{equation}
V(\phi)=\frac{\lambda}{4} M^4 \left[ 1+ B \, \frac{64\pi^2}{m_{{\rm
	Pl}}^4} \, \phi^4 \right] \,,
\label{eqn:fvqV} 
\end{equation}
where $B$ is a constant. For suitable $B$, this model is characterized by two 
separate epochs
of inflation.  For large $\phi$ values the false vacuum term is
negligible and the model becomes a $\phi^4$ potential, while for small
$\phi$ values the model becomes false vacuum dominated. Depending on
the value of the parameter $B$, these two epochs can be punctuated by
a brief suspension of inflation, while the potential becomes
temporarily too steep to support inflation. Upon reaching the lower part of the 
potential the field fast-rolls until slowed down by friction from the expansion.

\begin{figure}[t] 
\centering 
\leavevmode\epsfysize=6cm \epsfbox{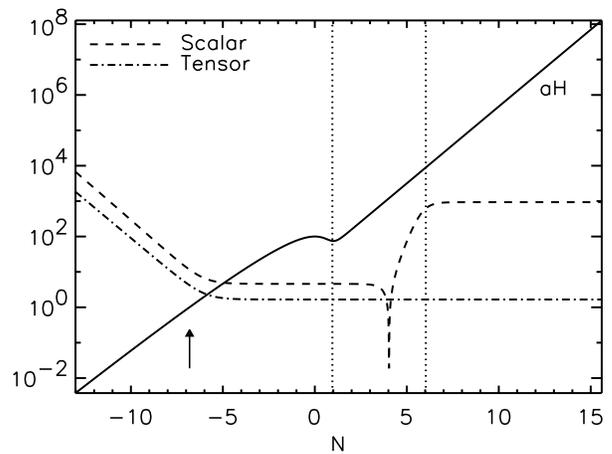}\\
\caption[fig1]{\label{fig:mode} The evolution of scalar and tensor
perturbations, $|{\mathcal R}_{{\rm k}}|$ and $|V_{{\rm k}}|$, for a
single mode $k=10^{-2}$ which is outside the horizon by a factor of 100
when inflation is suspended. The solid line is the quantity $aH/k$,
and $z'/z<0$ between the two vertical dotted lines. The arrow indicates
when $k=aH$, the instant of horizon crossing. The absolute
normalization of the perturbations is arbitrary, though the relative
one is correct.}
\end{figure}

The evolution of the quantity $1+\epsilon-\eta$ and the comoving Hubble
wavenumber, $aH$, is illustrated in Fig.~\ref{fig:aH} for the choice $B=0.55$. 
In this example
inflation is suspended for about $1$ $e$-fold, but $z'/z$ remains negative
for about 5 $e$-folds indicating that the field continues to fast-roll for some 
time after inflation restarts. It is during this latter period that scalar modes
which have recently left the horizon feel the effect of the driving term in 
Eq.~(\ref{eqn:mode2}).

We set the correspondence of scales such that $k=1$ corresponds to the
scale which equals the horizon at the time when inflation stops. This
correspondence is arbitrary, depending on the
mechanism for ending inflation and in particular the value of $\phi$
at which inflation ends.

The evolution of perturbations on a particular scale $k$, again for $B=0.55$,
is shown in Fig.~\ref{fig:mode}, obtained numerically using the approach of
Ref.~\cite{GL1}.  Even though this particular mode left the
horizon around $7$ $e$-folds before $z'/z$ turns negative, the
residual ${\mathcal R}_{{\rm k}}'$ given by
Eq.~(\ref{eqn:rdot_suppression}) is quickly blown up by the
exponentially growing driving term. After some time the exponential
growth becomes important and it drives $\left|{\mathcal R}_{{\rm
k}}\right|$ through zero and on to a much larger amplitude. Although
the mode is well outside the horizon at this time, its amplitude is
enhanced by a factor of around a hundred; we will discuss the physical 
interpretation of this further later. Once $z'/z$ becomes
positive again normal friction domination resumes, freezing out
$\left|{\mathcal R}_{{\rm k}}\right|$ at the enhanced amplitude. The
standard approximations therefore fail by a factor of around a hundred in
this case, which as we will see is by no means the worst. The tensor
amplitude is unchanged, so the tensor-to-scalar ratio is suppressed.

\begin{figure}[t]
\centering 
\leavevmode\epsfysize=6cm \epsfbox{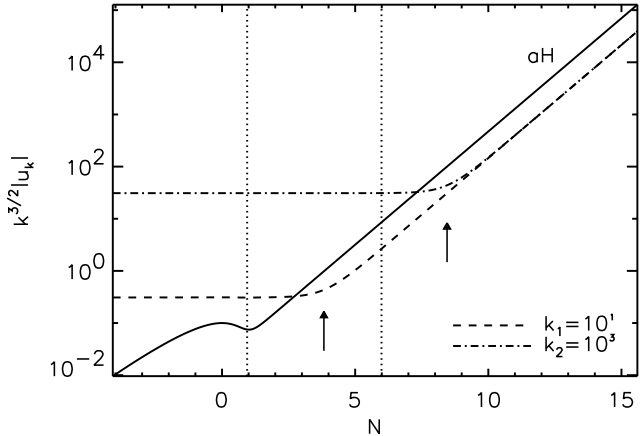}\\
\caption[fig1]{\label{fig:mode2} The evolution of two scalar modes, $|u_{{\rm 
k}}|$,
as in Fig.~\ref{fig:mode}. The arrows again indicate horizon crossing for each 
mode. For the first mode, $k_1$, horizon crossing
occurs while $z'/z<0$, while for the second it
occurs after $z'/z$ becomes positive again. Both modes asymptote to
approximately the same amplitude.}
\end{figure}

In Fig.~\ref{fig:mode2} we consider two other modes, this time
tracking the evolution of $|u_{{\rm k}}|$. One mode crosses
the horizon during the epoch when $z'/z<0$. Clearly Eq.~(\ref{eqn:mode1}) does 
not depend explicitly on $z'/z$,
and so the transition from the oscillating regime still occurs when $k^2=z''/z$.	
Scalar modes that leave the horizon after $z'/z$ turns
positive again, as is the case in the second mode shown in
Fig.~\ref{fig:mode2}, asymptote to a value that is independent of the
influence of $z'/z$ turning negative. This is to be expected, since sub-horizon
modes do not feel the influence of any background quantities.

\begin{figure}[t]
\centering 
\leavevmode\epsfysize=6cm \epsfbox{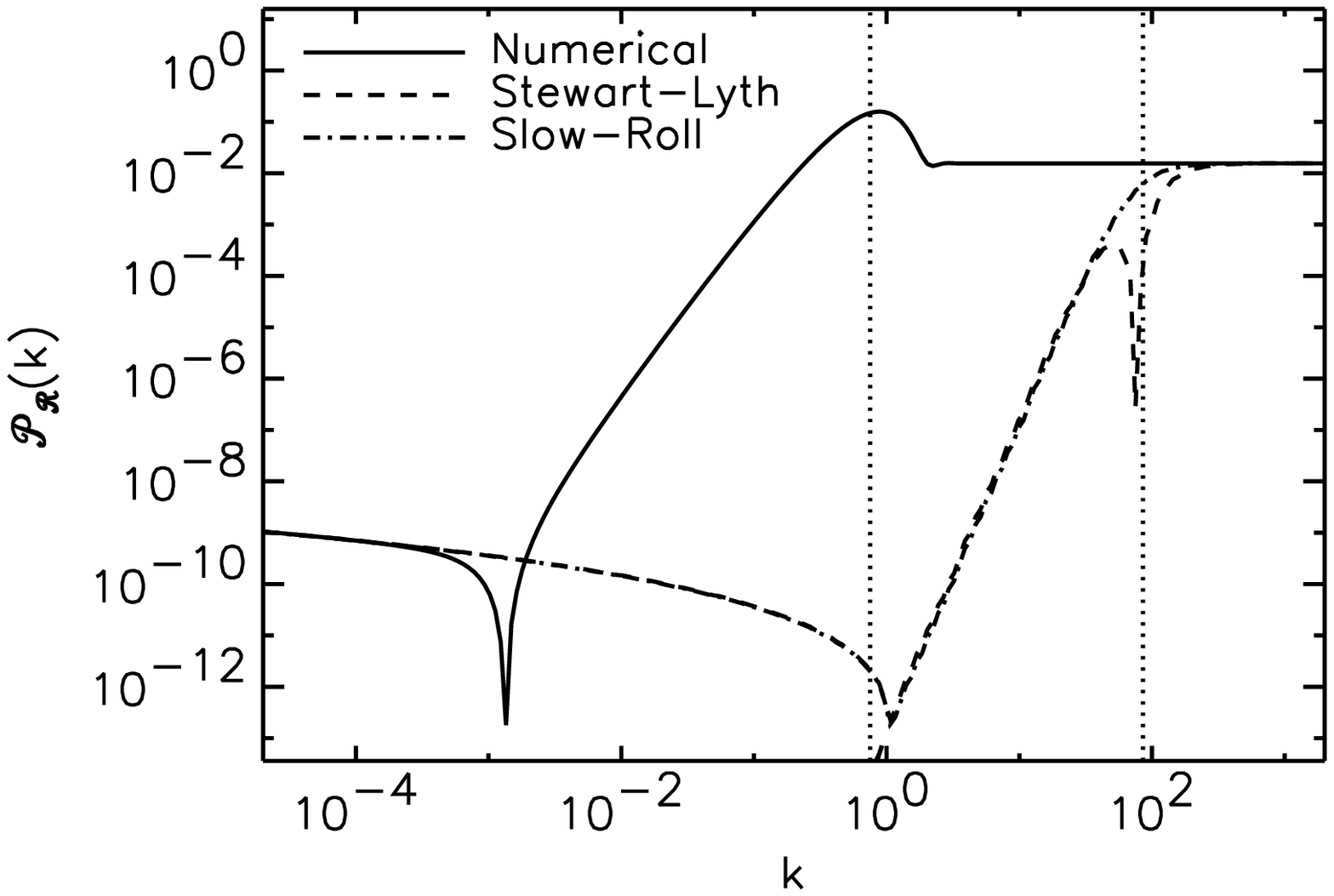}\\
\leavevmode\epsfysize=6cm \epsfbox{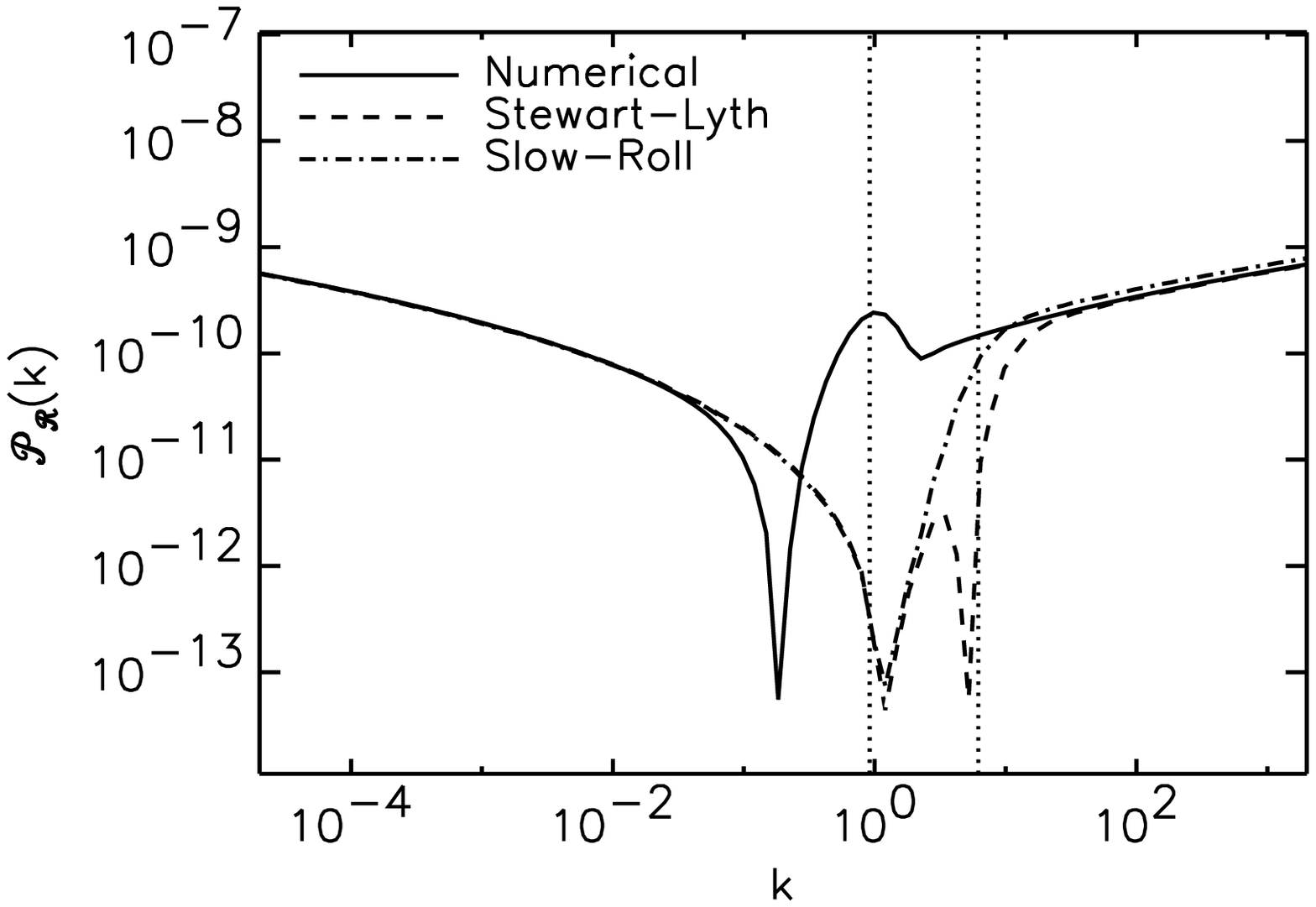}\\
\caption[fig1]{\label{fig:scal_spec} The scalar power spectrum as
determined from mode-by-mode integration. The upper panel shows $B=0.55$, while 
the lower is for $B=0.3$ (note that for inflation to be interrupted requires
$B > 0.19$ \cite{RLL}). The scales between the two
dotted lines correspond to the epoch when $z'/z<0$.  The first
discontinuity in the analytic expressions occurs in
the region where modes make multiple horizon crossings, while the
second sharp feature in the Stewart--Lyth case occurs due to an accidental 
cancellation of the entire Stewart--Lyth coefficient and has no physical
significance.}
\end{figure}

The overall result is that an extremely broad feature arises in the scalar power
spectrum.  This is shown in Fig.~\ref{fig:scal_spec}, where the spectrum for two
different choices of $B$ has been computed mode by mode, interpolating between
the two different epochs of inflation.  For comparison we plot the amplitude
predicted by the slow-roll and Stewart--Lyth formulae, where we use the
background values and slow-roll parameters derived from the numerical evolution.
For a small range of scales near $k=aH|_{{\rm end}}$ there is some ambiguity as
to when the formulae should be applied, since these modes leave the horizon
twice.  We plot both possible values.  The slow-roll and Stewart--Lyth
predictions fail by orders of magnitude for many $e$-folds (potentially enough
to encompass all scales accessible to large-scale structure observations in the
$B=0.55$ case), resuming their usual good agreement with the numerical results
soon after $z'/z$ turns positive again.

An intriguing feature of the spectrum for $B=0.55$ is the very flat portion on 
the right, which arises even though the disagreement with the slow-roll 
prediction indicates that we are nowhere near the slow-roll limit. In fact, this 
is a realization of the more general circumstance for a scale-invariant spectrum 
uncovered by Wands \cite{W} using duality arguments.
During this epoch, the field is fast-rolling along a relatively flat section of 
the potential, obeying
\begin{equation}
\ddot{\phi}+3H\dot{\phi}+V_{,\phi}\simeq \ddot{\phi}+3H\dot{\phi}=0 \,.
\end{equation}
The solution is $\dot{\phi}\propto a^{-3}$, giving $z\propto \tau^2$ rather than 
the $z \propto \tau^{-1}$ typical of slow-roll inflation. Nevertheless, this 
relation between
$z$ and $\tau$ leads to a scale-invariant spectrum of scalar perturbations 
\cite{W}, just as in the slow-roll limit. This decaying
solution of $\dot{\phi}$ is always present but is usually neglected.
Seto et al.~\cite{SYK} have shown that in extreme cases, when $\dot{\phi}$
becomes equal to zero, the slow-roll amplitude of the density perturbation as
written in Eq.~(\ref{eqn:scalpert}) will completely fail unless $\dot{\phi}$
is replaced by the corresponding slow-roll velocity $\dot{\phi}_{\rm s}=
-V_{,\phi}/3H$.

As the field slows down, one might have expected a feature to be produced. 
However, as long as $\epsilon$ is small, a necessary and sufficient
condition for a scale-invariant spectrum is that the entire square-bracketted 
term in Eq.~(\ref{eqn:d2zdt2}) remains constant in time,
which taking $\epsilon \ll 1$ reduces to 
\begin{equation}
\xi^2+\eta^2-3\eta\,=\,{\rm constant} \,.
\end{equation}

This encompasses both slow-roll ($|\eta| \ll 1$) and fast-roll ($\eta \gtrsim 
1$) inflation, as well as any smooth transition
between the two. This condition will be automatically satisfied as long as
the inflaton is effectively massless, allowing us to neglect the last term of
Eq.~(\ref{eqn:xi2}). However obtaining the flat portion required significant 
fine-tuning of $B$; notice that with $B=0.3$ the potential is much less flat 
in the corresponding region, though still flatter than the slow-roll formulae 
predict.

In the limit of an instantaneous transition from slow-roll to fast-roll
behaviour, inflation is no longer suspended and we arrive at Starobinsky's model
\cite{S}.  In this case the inflaton potential is characterized by a sudden
gradient discontinuity and the power spectrum takes on a similar step-like form,
but with superimposed oscillations on the upper plateau.

We now return to the physical interpretation of the change in 
the curvature perturbation on super-horizon scales. Although ordinarily a single 
scalar field is associated with purely adiabatic perturbations, it can in fact 
support entropy perturbations if its velocity perturbation does not obey a 
generalized adiabatic condition with respect to the field perturbation 
\cite{GWBM}. Under quite general circumstances, however, single field inflation 
does only generate adiabatic perturbations \cite{WMLL}, with the entropy 
perturbation ${\mathcal S}$ associated with a general scalar field perturbation 
being non-zero but becoming highly suppressed, ${\mathcal S} \sim e^{-2N}$, once 
the mode becomes frozen in upon leaving the horizon 
\cite{GWBM}. The constancy of the curvature perturbation on 
super-horizon scales in the absence of entropy perturbations holds under 
extremely general circumstances \cite{WMLL}.

Our situation is an exception to this. An entropy perturbation is the 
only source of super-Hubble growth \cite{WMLL,GWBM}, and during the phase where 
$z$ 
decreases the entropy perturbation grows until it becomes significant enough to 
source the curvature perturbation. The further above the Hubble 
length a scale is, the more the entropy has been suppressed and hence more time 
is needed for the entropy to become significant, 
so the effect does not extend up to arbitrarily large scales. After slow-roll 
resumes the entropy dies away 
and the perturbations become purely adiabatic, remaining so thereafter but 
retaining the shift in amplitude. The influence of the scalar field entropy can 
be studied using the definitions of Ref.~\cite{GWBM}, where the entropy is given 
by
\begin{equation}
{\mathcal S} = \frac{2V_{,\phi}}{3\dot{\phi}^2 \left( 3H\dot{\phi} + V_{,\phi}
	\right)} \, \left[ \dot{\phi} \left(\dot{\delta\phi} - \dot{\phi}A
	\right) - \ddot{\phi} \, \delta \phi \right] \,,
\end{equation}
where $A$ is the perturbation to the metric lapse function which is related to 
the curvature perturbation via the constraint equation 
$A= 4\pi \dot{\phi}^2 {\mathcal R}/m_{{\rm Pl}}^2 H^2$ \cite{GWBM}.
The entropy measures the failure 
of the perturbed velocity to match the generalized adiabatic condition with 
respect to the field perturbation itself. It allows Eq.~(\ref{eqn:mode2}) to be 
rewritten as two first-order equations
\begin{eqnarray}
\label{rpreq}
\frac{{\mathcal R}'}{aH} & = &  
	\frac{3}{2} \, \frac{3-2\eta}{3-\eta} \, {\mathcal S} \,; \\
\label{spreq}
\frac{{\mathcal S}'}{aH} & = & \left[ 
	\frac{3(\epsilon\eta-\xi^2)}{(3-2\eta)(3-\eta)}-3-\epsilon+2\eta 
	\right] \, {\mathcal S}\nonumber \\
 & & \hspace*{2cm} - \frac{2}{3} \, \frac{3-\eta}{3-2\eta} \,
  \frac{k^2}{a^2H^2}\, {\mathcal R}  \,.
\end{eqnarray}
Even in the slow-roll case, care is required in deriving the late-time evolution 
of the entropy, because its suppression is strong enough that the effect of the 
${\mathcal R}$ term does not become negligible despite its $k^2/a^2H^2$ 
prefactor. If we were able to neglect that term, and taking the slow-roll limit, 
one would find ${\mathcal S}'/aH \simeq -3 {\mathcal S}$ implying ${\mathcal S} 
\sim e^{-3N}$, but we see that this does fall off faster than the last term. The 
self-consistent solution therefore has late-time behavior ${\mathcal S} \sim 
e^{-2N}$, and indeed is what we see in our numerical simulations.

During fast-roll the situation is very different, because ${\mathcal S}$ is able 
to grow and so the influence of the final term in Eq.~(\ref{spreq}) becomes 
negligible. Making the false vacuum
($\epsilon \ll 1$) and massless ($V_{,\phi\phi} \ll H^2$) assumptions, 
Eq.~(\ref{spreq}) can then be 
written as
\begin{equation}
\frac{{\mathcal S}'}{aH} \simeq \left[ 2\eta -3 + \frac{3\eta}{2\eta-3} 
	\right] {\mathcal S} \,,
\end{equation}
During fast-roll $\eta \simeq 3$, and the entropy
can swiftly grow $\sim e^{6N}$ until it becomes large enough to have a 
significant impact on ${\mathcal R}$. Once fast-roll ends the $k^2{\mathcal R}$
term in Eq.~(\ref{spreq}) is initially small giving the ${\mathcal S} \sim 
e^{-3N}$ behaviour, but soon after the curvature terms re-asserts itself 
restoring the ${\mathcal S} \sim e^{-2N}$ late-time behaviour. However this 
transition is not of any particular physical significance since the curvature 
perturbation itself has long since approached its asymptotic value.

One might wonder whether the entropy perturbations in the scalar field
could survive as such after inflation, but that does not seem to be possible.
Either fast-roll is followed by slow-roll as just discussed, or it is followed
by inflation ending through decay of the inflaton.  That decay will lead to a
purely radiation-dominated universe, which is unable to support non-adiabatic
perturbations. The process of decay of the inflaton would therefore remove the
entropy source term.  Note also that considerable fine-tuning is required to
stay in the fast-roll regime for a prolonged period; in order to 
have inflation the initial kinetic energy can at most be of order of the 
potential energy, but then the kinetic
energy falls off very quickly, and yet for fast-roll to be sustained its
contribution to the scalar wave equation must remain larger than that from the
slope of the potential.


\section{Perturbations at the end of inflation}

A second scenario where the standard equations cannot be directly
applied is at the true end of inflation. This is an important regime
because such short-scale perturbations can lead to the formation of
primordial black holes, which gives the most important constraint on the
late stages of inflation.  There are two standard mechanisms for
ending inflation, one being the hybrid method of an instability in
another direction in field space and the second being the breakdown of
slow-roll (see Ref.~\cite{LLbook} for a review). A particularly relevant case is 
in hybrid inflation models
such as the running-mass model \cite{running}, where the slow-roll
approximation is only well respected over a limited range of scales.
In the running-mass model slow-roll inflation comes to an end due to
$\eta$ growing, but inflation may continue in the fast-roll regime until
an instability is reached.

Typically the standard formulae for the perturbations will break down
near the end of inflation, because of a failure to reach the
asymptotic limit (see e.g.~Ref.~\cite{HLZ}) and often
because the slow-roll approximation is not accurate. For example,
in the running-mass model Eq.~(\ref{eqn:z-flip}) becomes
\begin{equation}
1-\eta \simeq 0 \,,
\label{eqn:z-flipapprox}
\end{equation}
and so $\eta=1$ is the last point at which constraints based on the
shape and size of a blue inflationary power spectrum can be reliably
applied without resorting to mode-by-mode integration
of the power spectrum~\cite{LGL}.  

We are not able to study the hybrid case due to the complexity of the
multi-field dynamics (though see Ref.~\cite{GWBM} for a general formalism
for doing so), so we restrict our attention to single-field
models ending by violation of slow-roll. This is not in fact the most
interesting case as typically such models have a red spectrum where
small-scale perturbations are not very significant, but illustrates
the main physics. For simplicity we study the quadratic chaotic
inflation model $V(\phi) \propto \phi^2$.

\begin{figure}[t]
\centering 
\leavevmode\epsfysize=6cm \epsfbox{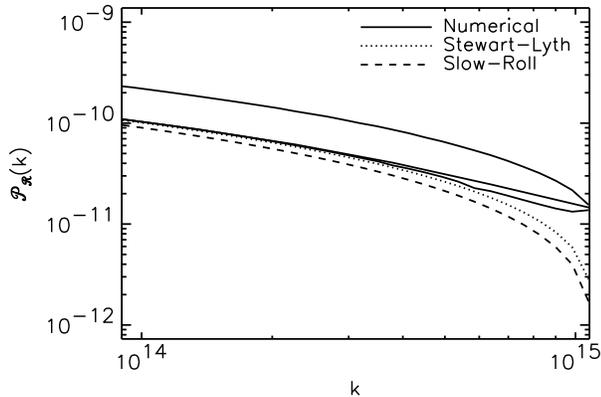}\\
\caption[fig1]{\label{fig:scal_spec2} The scalar power spectrum plotted at 
various stages of its evolution. The numerical spectra are evaluated (from top 
to bottom) at horizon exit, at the end of inflation, and at horizon re-entry for 
each mode. The right-hand edge of the plot corresponds to the wavenumber 
equalling the Hubble radius at the end of inflation. The Stewart--Lyth and 
slow-roll spectra are evaluated, as usual, at horizon exit. } 
\end{figure}

In Fig.~\ref{fig:scal_spec2} we plot the scalar power spectrum for the
last few $e$-folds of modes to leave the horizon before inflation
ends. Because the asymptotic regime is not adequately reached at any stage, 
there is no preferred choice as to when to plot the amplitude, and we evaluate 
it at three stages: horizon crossing during inflation, at the end of inflation, 
and at horizon re-entry after inflation once the background field begins 
oscillating. Notice that the amplitude at horizon exit is typically much greater 
than the value at the end of inflation, and that the slow-roll and Stewart--Lyth 
approximations indeed resemble the latter rather than the former.

We see that the slow-roll and Stewart--Lyth formulae underpredict the
asymptotic and re-entry amplitudes of perturbations which exited the
horizon close to the end of inflation.  One therefore expects an enhancement of
primordial black hole production, though a detailed calculation would need to
track the derivative of the curvature perturbation at re-entry as well as its
amplitude.  This result suggests that normally quoted constraints are on the
conservative side, though typically the correction would not be large.  We
mention additionally that for perturbations which do not reach the asymptotic
regime there are questions as to how the quantum-to-classical transition might
take place (see e.g.~Ref.~\cite{Q2C}); we will not however attempt to address
this here.

\section{Discussion}

The accuracy of the usual analytic expressions for the inflationary power 
spectrum depends on scales evolving smoothly through horizon crossing and into 
the asymptotic regime $k^2 \ll z''/z$. In this paper, we have investigated two 
situations where this is not achieved, one being a temporary end to inflation 
and the other the true end. In the former case we have seen that the modes can 
have a very complicated evolution, including the possibility of amplification on 
super-horizon scales via an exponential driving term corresponding to an
entropy perturbation in the scalar field. Such behaviour can be 
traced to the complicated evolution of the scalar pump field $z$, which no 
longer grows monotonically. The net effect is the insertion of a broad feature 
into the power spectrum that can only be computed by mode-by-mode integration, 
and which can differ wildly from the slow-roll and Stewart--Lyth approximations. 
Typically the features are sufficiently non-scale-invariant to be excluded 
already on astrophysical scales, but we have also seen 
that a very flat spectrum can be obtained while far from slow-roll, confirming 
an analytic analysis by Wands~\cite{W}.

The second scenario we studied was the true end of inflation, where the failure 
to reach an asymptotic regime means that perturbations re-enter the horizon with 
a higher amplitude. The most interesting physical consequence of such modes is 
in primordial black hole formation, and this result indicates that earlier 
treatments assuming the slow-roll formulae are somewhat conservative, as the 
mode equation solutions indicate that these formulae underestimate the 
perturbation amplitude at horizon re-entry.

\section*{Acknowledgments}

S.M.L.~is supported by PPARC. We thank Ian Grivell for making his mode
function code available to us, and Cyril Cartier, Misao Sasaki, Alexei 
Starobinsky and David Wands for useful conversations. We acknowledge the
use of the Starlink computer system at the University of Sussex,
and thank the Observatoire Midi-Pyr\'en\'ees for hospitality while
part of this work was carried out.

 
\end{document}